%% file: main.tex
\documentclass[twocolumn, secnumarabic,  floatfix, nofootinbib, nobalancelastpage, longbibliography]{revtex4-2}

\usepackage[bottom]{footmisc}

\def\preprintlinklocation{http://dft.uci.edu}

\input Burke_template
\input KieronsMacros

\graphicspath{{f/}}

\def\sss{\scriptscriptstyle\rm}

\def\ben{\begin{equation}}
\def\een{\end{equation}}

\def\HF{^{\rm HF}}
\def\conv{^{\rm conv.}}

\def\c{_{\sss C}}
\def\s{_{\sss S}}
\def\br{{\bf r}}

\begin{document}
\sf\small
\coloredtitle{
Can the Hartree-Fock kinetic energy exceed the exact kinetic energy?
}
\coloredauthor{Steven Crisostomo}
\affiliation{Department of Physics and Astronomy, University of California, Irvine, CA 92697}

\coloredauthor{Mel Levy}
\affiliation{Department of Chemistry and Quantum Theory Group, Tulane University, New Orleans, LA 70118}

\coloredauthor{Kieron Burke}
\affiliation{Department of Chemistry, University of California, Irvine, CA 92697}
\affiliation{Department of Physics and Astronomy, University of California, Irvine, CA 92697}

\date{\today}

\begin{abstract}
The Hartree-Fock (HF) approximation has been an important tool for quantum-chemical calculations since its earliest appearance in the late 1920s,
and remains the starting point of most single-reference methods in use today.
Intuition suggests that the HF kinetic energy should not exceed the
exact kinetic energy, but no proof of this conjecture exists, despite a near century of development. Beginning from a generalized virial theorem derived from scaling considerations, we derive a general expression for
the kinetic energy difference that applies to all systems. 
For any atom or ion this trivially reduces
to the well-known result that the total energy is the negative of the kinetic
energy and since correlation energies are never positive, proves the
conjecture in this case.   Similar considerations apply to molecules
at their equilibrium bond lengths.
We use highly precise calculations on
Hooke's atom (two electrons in a parabolic well)
to test the conjecture in a non-trivial case, and to parameterize
the difference between density-functional and HF quantities,
but find no violations of the conjecture.
\end{abstract}

\maketitle

\sec{\label{sec:intro} Introduction \& Motivation}
This paper addresses the simple question: In a typical electronic structure problem, can the Hartree-Fock \cite{H28,S28,F30} (HF) kinetic
energy ever exceed the exact kinetic energy?  And if not, can we 
prove that it cannot? This is an intriguingly simple question of principle
that is surprisingly unaddressed, despite the use of HF as a starting
point for many modern quantum-chemical methods, such as MP2 \cite{C11} or CCSD(T) \cite{K02}, or Green's function calculations in materials, such as GW \cite{He99}. Perhaps even more surprising is that, within density functional theory (DFT),
where the definition of correlation is subtly different, it has long
been known that the Kohn-Sham (KS) kinetic energy can never exceed the
exact kinetic energy. The simplicity of the proof in this case
relies on both quantities being defined as density functionals, so the
comparison is made on the same density \cite{KS65,kb99}. In HF theory, the HF density
differs from the exact density, which complicates matters considerably.

Here, we limit consideration to spin-unpolarized
non-relativistic, non-magnetic electronic
calculations. Moreover, we consider only those cases where symmetries
are not broken, and the HF minimizer is a single Slater determinant.
Nonetheless, we are unable to show that the HF kinetic energy cannot
exceed the exact kinetic energy. 

There are two key differences between HF quantities and KS quantities.
While their definition in terms of orbitals is identical, the former
is evaluated on those orbitals that minimize the HF energy for a given
potential, while the latter are those orbitals that minimize only
the kinetic energy on a given density. In the special case
of just two electrons, the one occupied orbital is trivially determined
by the density, so only the difference in densities remains.  Even in 
this simpler case, we are unable to prove our conjecture. However, we do derive a virial relation for the HF kinetic energy
difference that applies to all cases. Our result shows that
the difference vanishes as the error in the HF density vanishes, i.e.,
in the high-density (or weakly correlated) limit.  

In the absence of a proof, the next best thing is to look for 
a counterexample. This is less trivial than it at first appears.
For atoms and ions, and for molecules at equilibrium, it is trivial
to show that, because correlation energies are negative, the HF
kinetic energy is never greater than the exact value.   
Moreover the differences, especially those with DFT values, where the
conjecture is true, are extremely small, making them 
very difficult to determine with sufficient accuracy using standard atom-centered basis sets. 

Instead we turned to the two-electron Hooke's atom. The quadratic
external potential ensures that the conjecture is not trivially
true, allowing the possibility of a counter example. Very precise numerical calculations demonstrate that the conjecture
is correct for all values of the oscillator frequency for which the 
restricted HF solution is valid.

Why might this question be of practical importance? Exact conditions,
such as the positivity of the kinetic correlation energy, are used
in density functional theory all the time \cite{WCh12}. Relationships between the total energy and kinetic energy (exact and approximate) have aided in the development of correlated basis sets for quantum-chemical calculations \cite{H85} . Exact conditions are built in to 
approximations, or their violation is checked in approximations
that do not automatically satisfy such conditions.   Apart from 
intellectual curiosity, it may prove useful in the future of 
wavefunction theory to prove analogs. Although here we have
failed, we hope our first steps might allow others to succeed.

A practical problem of recent interest where this is very relevant
is in the use of HF densities as inputs to hybrid density functionals,
which depend on the KS orbitals \cite{GJPF92,SVSB22}. In principle, one should
perform a KS inversion to extract the KS orbitals for the HF density,
but in practice it is infinitely easier to use the HF orbitals themselves. This introduces small differences in the kinetic energy and the vital question is:  Are these small relative to the differences produced by the density?  The answer appears to be yes they are, but
accuracy is limited by the limitations of KS inversions in atom-centered
basis sets \cite{NS21,NM21}.
 
Recently it has been shown that there are certain regions of the Hubbard dimer where the HF kinetic energy becomes greater than the exact value \cite{GPJ22}. The authors report that the Hubbard dimer HF kinetic energy exceeds the exact value approximately when the on-site interaction strength is comparable to the external potential difference between the sites. Our conjecture only pertains to real space Hamiltonians, unlike the case of the dimer \cite{CF49,CB15}.  Discrete lattice systems, such as the Hubbard model and its variants, require separate considerations as many basic theorems do not hold as they would in real space \cite{PvL22}.

\sec{\label{sec:Theory}Theory}

Consider a Hamiltonian of the form
\ben \hat{H} = \hat{T} + \hat{V}_{ee} + \hat{V}, \een
where $\hat{T}$ is the kinetic energy, $\hat{V}_{ee}$ is the electron-electron interaction, $\hat{V}$ is a one-body multiplicative external potential  $\hat{V} = \sum_{j=1}^{N}v(\br_{j})$, with atomic units used throughout. Consider a scaled wavefunction $\Psi_{\gamma} = \gamma^{3N/2}\Psi(\gamma \br_{1},...,\gamma\br_{N})$, whose $\gamma =1$ value corresponds to the exact groundstate wavefunction of $\hat{H}$. From the variational principle, the scaled wavefunction $\Psi_{\gamma}$ must obey the following minimization condition
\ben \frac{d}{d \gamma} \bra{\Psi_{\gamma}}\hat{H}\ket{\Psi_{\gamma}}\bigg\rvert_{\gamma =1} = 0. \een
Using scaling properties of the energy operators \cite{HK37, L59}, the minimization condition can be expressed as 
\ben
\frac{d}{d\gamma}\Big\{\gamma^{2} T + \gamma V_{ee} + \int d^{3}r \   v(\br)  n_{\gamma}(\br)\Big\}\bigg\rvert_{\gamma =1}= 0, \label{eq:sca1}
\een
where $n_{\gamma}(\br) = \gamma^{3}n_{v}(\gamma \br)$ is the scaled groundstate one-body density. 
Computing the derivative and rearranging yields the virial theorem for the groundstate energy:
\ben
E = -T + I[v,n], \label{eq:vir1}
\een
where
\ben
I[v,n] = \int d^{3}r \  \Big(1 + \br \cdot \nabla \Big)v(\br)\, n(\br) ,
\een
is the virial of the potential $v(\br)$ with the groundstate density $n(\br)$.
The virial theorem holds for all potentials $v(\br)$ whose groundstate density $n(\br)$ vanishes sufficiently
rapidly at infinity.

An analogous virial relation to Eq.~(\ref{eq:vir1}) holds for the Hartree-Fock quantities because
the HF solution is a variational minimum and the energy scales identically to the exact solution.
Thus
\ben
E\HF = -T\HF + I[v,n\HF],  \label{eq:vir2}
\een
where $ n\HF(\br)$ is the HF groundstate density, i.e., 
the density produced by the single Slater determinant $\Phi\HF$ that minimizes the expectation value of the Hamiltonian $\hat{H}$. 
Now $E\c\conv =E-E\HF$ is the  conventional  quantum-chemical definition of the
correlation energy, originally credited to L{\"o}wdin \cite{L55_corr}. To distinguish L{\"o}wdin's definition from the DFT correlation energy, we refer to $E\c\conv$ simply as the conventional correlation energy.
We define the kinetic contribution to the conventional correlation energy via
\ben
T\c\conv = T-T\HF,
\een
and our conjecture is that this quantity is non-negative.
Taking the difference of Eq.~(\ref{eq:vir1}) and Eq.~(\ref{eq:vir2}) yields 
\ben
T\c\conv + E\c\conv = I[v,n-n\HF] \label{eq:vir3}.
\een

Equation (\ref{eq:vir3}) can be considered from several different perspectives.
Since $E\c\conv$ is non-positive (by the variational theorem), then only if $I$ becomes more
negative than $E\c\conv$ will our conjecture be violated.   However, we do not know if
$I$ in Eq.~(\ref{eq:vir3}) always has a definite sign.  We can expect it to involve tremendous
cancellations inside the real-space integral, as correlation energies are small, and kinetic
correlation energies are of the same order, while the inputs to the virial integral are not.

Next we list several special cases where $I[v,n-n\HF]$ vanishes.   First, in the limit
of weak correlation (such as the large-$Z$ limit of non-degenerate ions of fixed $N$ \cite{LS77,L81,C83}),
then $n\HF(\br)\to n(\br)$ sufficiently fast to make $I$ on the right vanish, while
$E\c\conv$ remains finite.   Thus, in such a limit, $T\c\conv\to |E\c\conv|$, guaranteeing its
non-negativity.   Second, for any potential $v(\br)=-Z/r$, such as an atom or
ion,  the virial itself identically vanishes, no matter what density it is evaluated on,
and the same conclusion can be drawn.  Finally, for any molecule at equilibrium (but not
otherwise), treating the nuclei as classical \cite{C60}, the virial for the electronic energy matches
the atomic result, yielding the same conclusion.  

The situation is eerily similar, but importantly different, for Kohn-Sham DFT.
In this case, all energies are defined as density functionals and are therefore
evaluated on the exact density for a given system.   For any given density,
the KS kinetic energy is defined to be the minimum over all possible wavefunctions:
\ben
T\s[n] = 
\min_{\Psi\to n}\bra{\Psi} \hat{T} \ket{\Psi}.
\een
Typically, the minimizer is a single Slater determinant, i.e., the KS wavefunction of KS orbitals.
Since $T$ is the exact interacting kinetic energy, by construction
\ben
T\c = T - T\s \geq 0,
\een
for any density, including the exact density of our system.
The scaling behavior of the correlation energy results in an analogous virial theorem \cite{LP85}
\ben
E\c + T\c = -J[v\c,n],
\een
where $v\c(\br)$ is the exact correlation potential and 
\ben
J[v,n] = 
\int d^{3}r \  \br \cdot \nabla v(\br)\, n(\br),
\een
is just the virial of the potential.  As we argue (and show) below, we expect this expression
to be quantitatively close to its HF analog, Eq.~(\ref{eq:vir3}), but its smallness arises in a very
different way.  Here, the potential is already small, but there is no density difference.
Finally, we note that, in the few cases with sufficiently precise evaluations, $J$ has never
been found negative, i.e., $T\c < |E\c|$, but this has also never been proven to be generally true.

Like the HF case, in the weakly interacting limit, the virial term vanishes, and
$E\c \to -T\c$.  Unlike the HF case, for atoms and ions and molecules at
equilibrium, the virial does not require $J$ to vanish, and in general it does not.
Thus $E\c$ is -42.1  \text{mH} for the He atom, but $T\c$ is 36.6  \text{mH}  \cite{HU97}.

Lastly, we discuss differences between HF and KS-DFT quantities.
First, we note that
\ben
\Delta E\c = E\c-E\c\conv \leq 0,
\een
because the HF energy minimizes the expectation value of the Hamiltonian over all orbitals, while the KS
scheme restricts orbitals to those from a single multiplicative KS potential \cite{GPG96}.
But typically $\Delta E\c$ is far smaller in magnitude than $E\c$.  In the
weakly interacting limit, $E\c\conv \to E\c$, so that $\Delta E\c$ vanishes.
For the He atom, $\Delta E\c$ is only -63 microHartree \cite{GPG96}.
The equivalent kinetic quantity is
\ben
\Delta T\c = T\c - T\c\conv = T\HF - T\s,
\een
which also vanishes in the weakly interacting limit.  Finally, we might expect in general
$\Delta E\c + \Delta T\c$ to be even smaller in magnitude than any object discussed so far.
However, this is not true when the virial makes $T\c\conv=-E\c\conv$, such as for the He atom.
In that case, $\Delta T\c$ is -5.5  \text{mH}, the difference being entirely due to the difference
between the exact and approximate densities.

In the next section, we give exact results for a two-electron system for which the virial of
the potential does not vanish.
\sec{Hooke's Atom}
\label{sec:hooke}

\begin{figure}[!htb]
    \includegraphics[width=\linewidth]{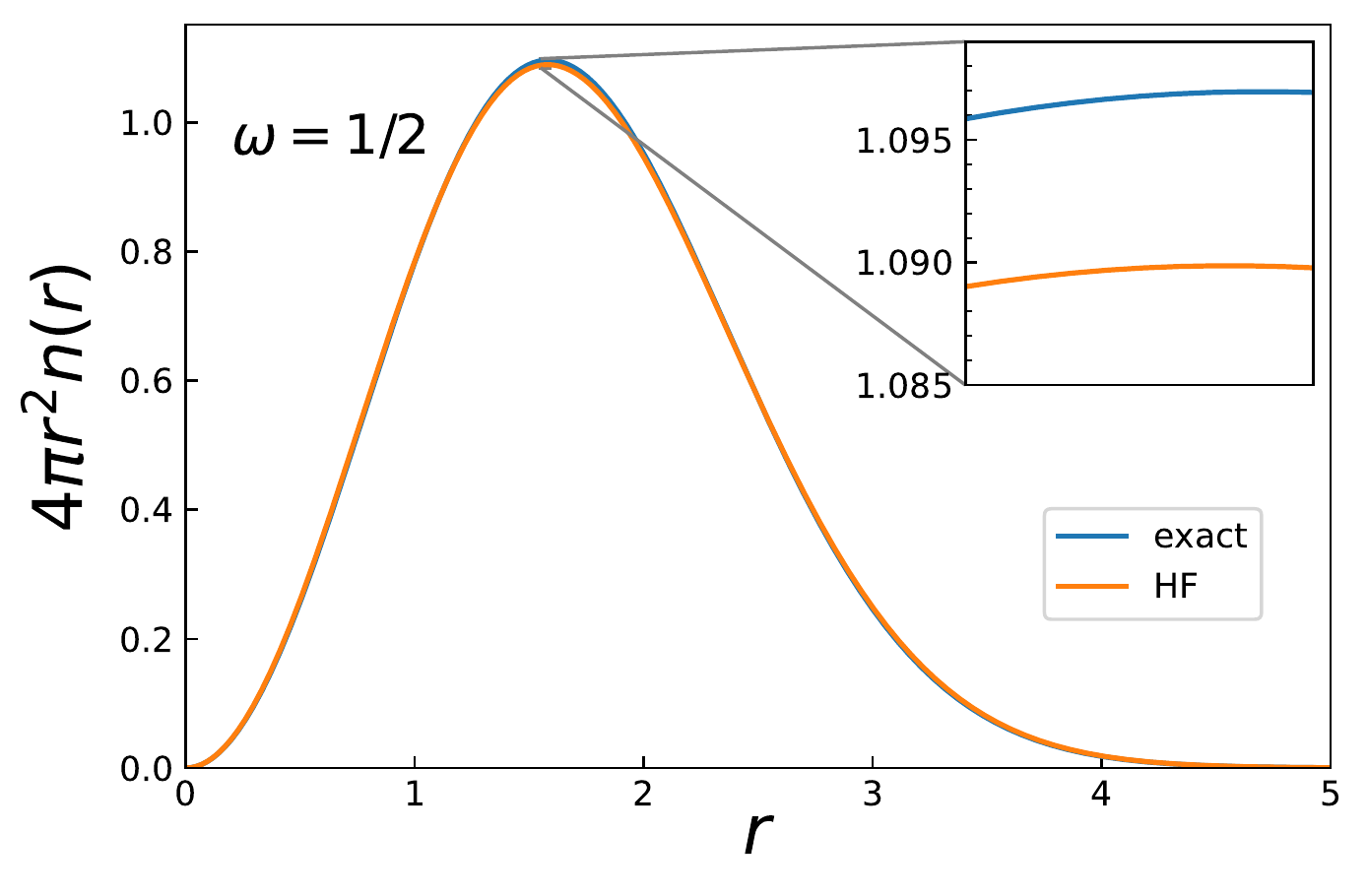}
    \label{fig:1a}
  \hfill 
    \includegraphics[width=\linewidth]{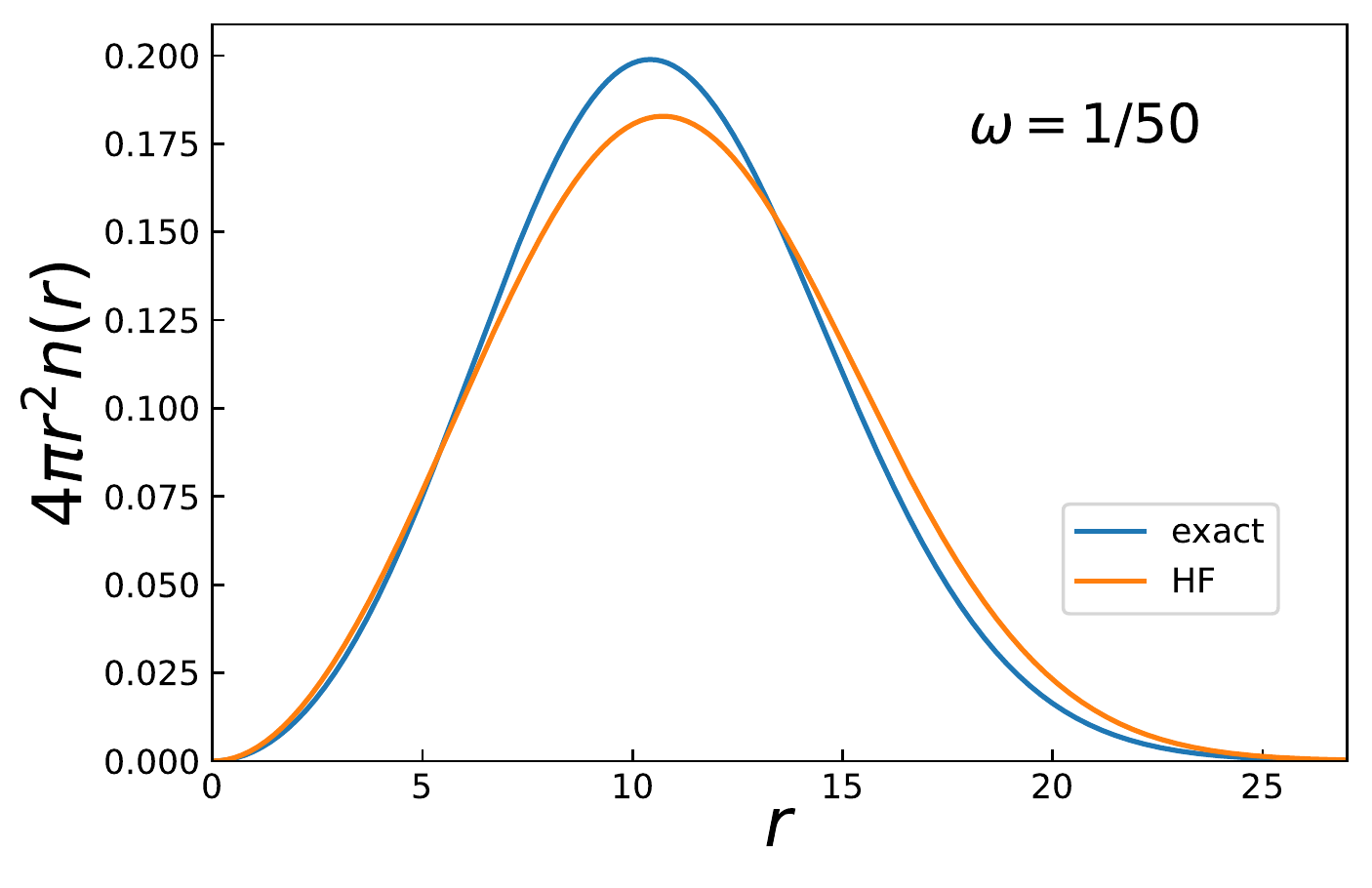}
    \label{fig:1b}
	\caption{Radial groundstate densities for moderate and small oscillator frequencies of Hooke's atom.}
	\label{fig:1}
\end{figure}

Hooke's atom consists of two electrons with Coulomb repulsion in a parabolic potential well $v(\br) = \omega^{2}r^{2}/2$. The presence of a parabolic potential does not permit a simplification of the virial result into the familiar atomic expression and it remains to be seen whether $T\c\conv$ is positive. Closed form solutions exist for certain values of $\omega$ \cite{KS62He,T93}, but for arbitrary values the solutions must be found numerically. The groundstate can be expressed formally as an infinite series which truncates whenever a closed form solution exists \cite{IBL99a}. No closed form expression exists for the Hartree-Fock density, and it must be found numerically as well, but the result converges rapidly when using Gaussian basis functions. We produce the relevant Hartree-Fock quantities using the basis functions of O'Neill and Gill \cite{OG03}. When plotted, the exact and HF densities coincide closely for large $\omega$ (as seen in Fig.~\ref{fig:1}), but their difference is vital to our question.


%
The following is an analysis of the kinetic energies and the terms which appear in the virial expression for the 
conventional correlation energy Eq.~(\ref{eq:vir3}). In the high-density $(\omega \to \infty)$ limit, the asymptotic behavior of the exact and HF kinetic energies are derivable from second order perturbation theory \cite{CP00,GO05}.

\ben E(\omega) = 3\omega + \sqrt{\frac{2\omega}{\pi}} - c + \mathcal{O}\Big(\frac{1}{\sqrt{\omega}}\Big), \label{eq:hdE}\een
where all energies are taken to be in Hartree units and the exact value of the constant is given by
\ben c = \frac{2}{\pi}\Big[1+\ln 2 \Big]-1 \approx  77.9\   \text{mH} . \een
For the Hartree-Fock problem the constant is exactly
\ben c\HF = \frac{4}{\pi} \Big[1 + \ln(8-4\sqrt{3})\Big] -\frac{1}{3} \approx 28.2\   \text{mH} . \een
The Hellmann-Feynman theorem \cite{F39,S62,CP00} yields
\ben I(\omega) = E(\omega) + T(\omega) =   \frac{3\omega}{2} \frac{d E(\omega)}{d \omega}, \een
and applies exactly and within HF \cite{YH79}.
Taking the difference between the exact and HF quantities, to $\mathcal{O}(\omega^{-1/2})$,
\ben \Delta I(\omega) \approx -\frac{d}{\sqrt{\omega}} \label{eq:HDI} , \een
where $\Delta I = I[v,n-n\HF]$. The constant $d$ has a closed form expression and is determined from the perturbation theory of the exact and HF energies. The numerical value can be found to many digits and is approximately $d \approx 7.03 \  \text{mH} $ \cite{WB70}. In the high-density limit we get the expected result that $E\c\conv \approx -T\c\conv$, and by extension $T>T\HF$. 

The low-density $(\omega \to 0)$ behavior of the total energy can be found by expanding the external potential about its classical equilibrium position and then computing the perturbation theory. At order $\mathcal{O}(\omega^{4/3})$ the energy is approximately  \cite{CP00}
\ben E(\omega) \approx 3\Big(\frac{\omega}{4}\Big)^{2/3} + 2(3+\sqrt{3})\Big(\frac{\omega}{4}\Big) + \frac{7}{9} \Big(\frac{\omega}{4}\Big)^{4/3}, \label{eq:ldE} \een
 with analogous expressions for $T$ and $I$.

Determining the small $\omega$ dependence of $T\HF$ and $I[v,n\HF]$ has several complications. Throughout all calculations we assumed that the Hartree-Fock wavefunction remains a single Slater determinant, but in the strongly correlated limit, the minimizing Hartree-Fock wavefunction might be spin unrestricted or multi-determinantal in nature. For small values of $\omega$ the HF eigenvalue problem becomes numerically unstable, but we can show that $E\c\conv$ closely agrees with the form of Eq.~(\ref{eq:ldE}).
\ben
E\c\conv(\omega)  \approx -\omega^{2/3}\sum_{j=0}^{5}a_{j}\omega^{j/3} \quad \quad \quad (\omega \to 0),
\label{eq:ldf}\een
and a similar form for $-T\c\conv$, where all coefficients have been determined numerically. According to the perturbation theory in the high-density limit, $E\c\conv$ can be closely approximated by a series in powers of $\omega^{-1/2}$.

\ben
E\c\conv(\omega) \approx -\sum_{j=0}^{4}(-1)^{j}b_{j}\omega^{-j/2} \quad \quad \quad (\omega \to \infty),
\label{eq:hdf}\een
again, with a similar form for $-T\c\conv$ and where the coefficients are determined numerically. The coefficients for $T\c\conv$ and $\Delta I$ were determined from the fit of $E\c\conv$ in both the low and high-density limits via the Hellmann-Feynman theorem.

\begin{figure}[htb]
\begin{center}
\includegraphics[angle=0,width=8.5cm]{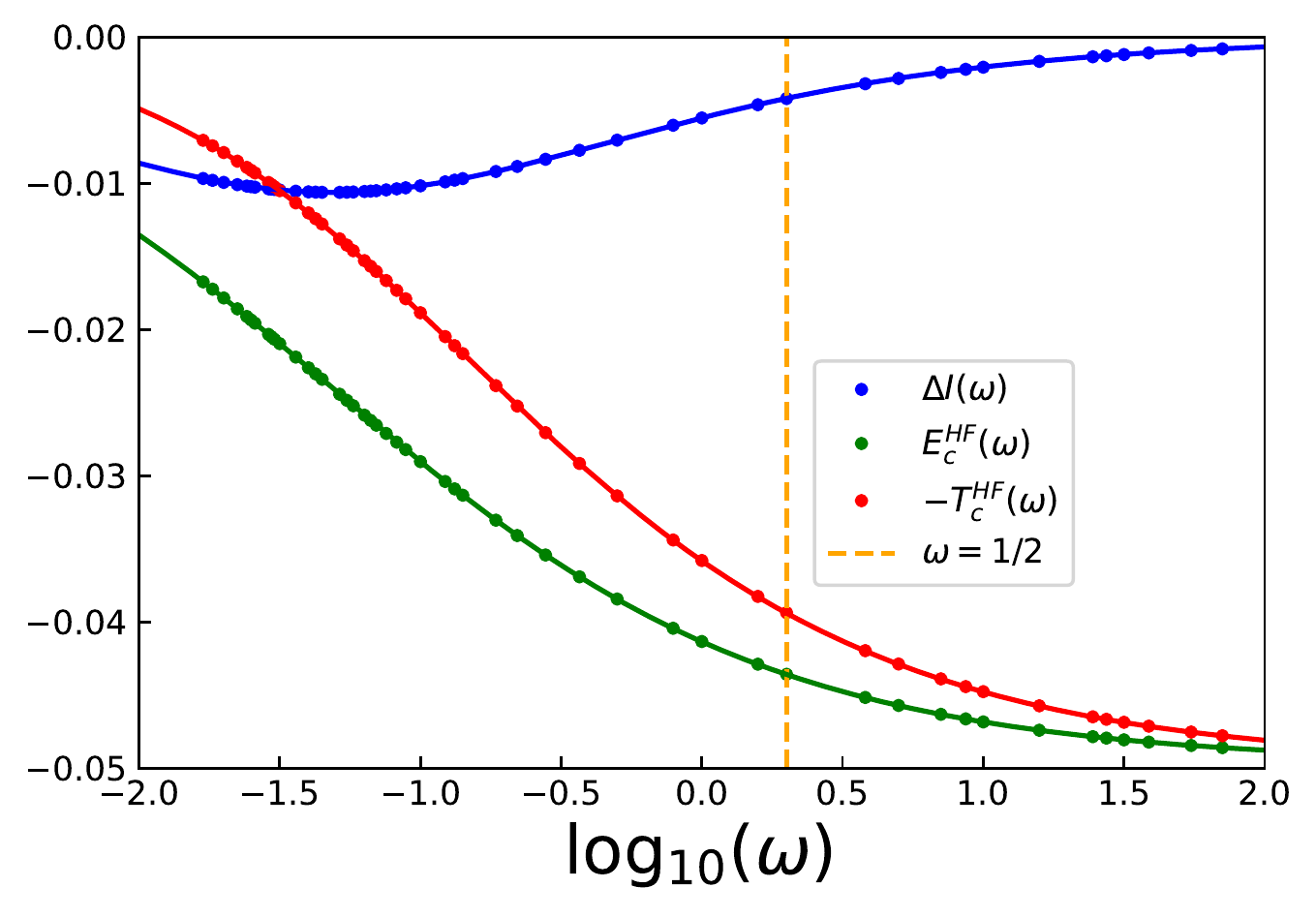}
\end{center}
\caption{Calculated correlation energies (points) and their approximate fits (solid) in Hartrees.}
\label{fig:2}
\end{figure}

\begin{figure}[!htb]
\includegraphics[width=\linewidth]{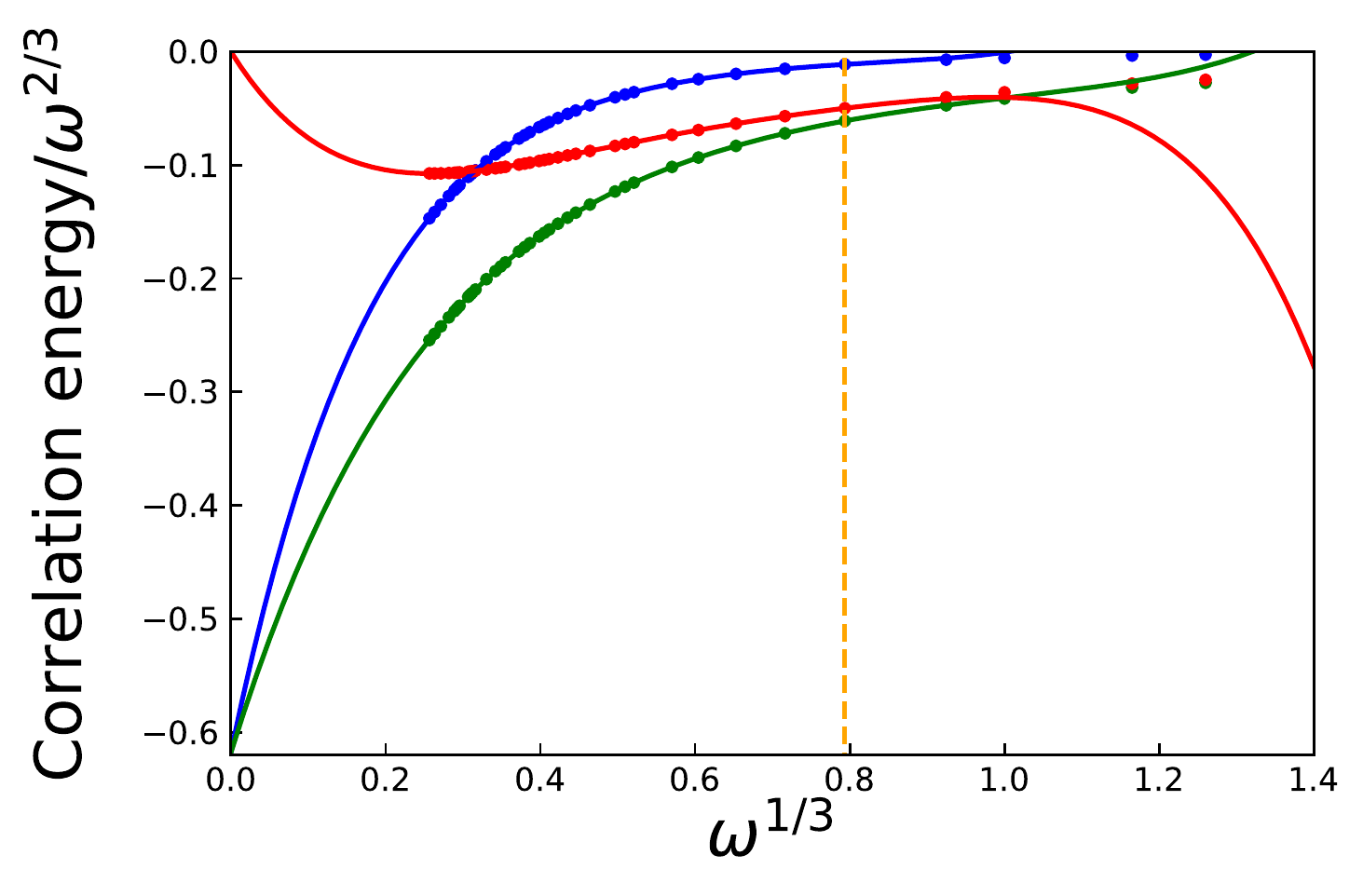}
\label{fig:3a}
\hfill 
\includegraphics[width=\linewidth]{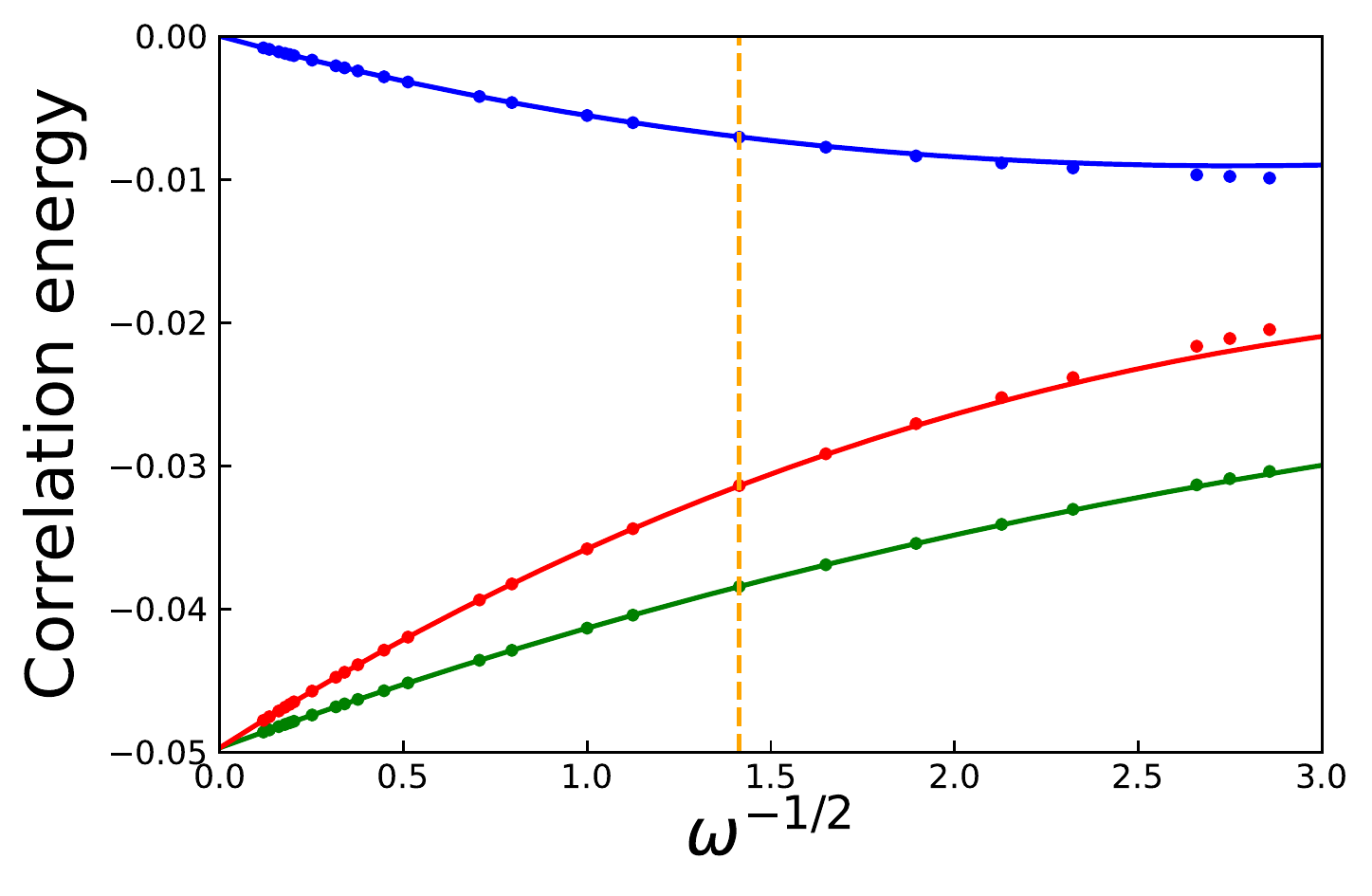}
\label{fig:3b}
\caption{Energies and fits in low (top) and high (bottom) density limits, in Hartrees: $E\c\conv$ (green), $-T\c\conv$ (red), $\Delta I$ (blue) and dashed orange line denotes $\omega =1/2$.}
\label{fig:3}
\end{figure}

\begin{figure}
\begin{center}
\includegraphics[angle=0,width=8.5cm]{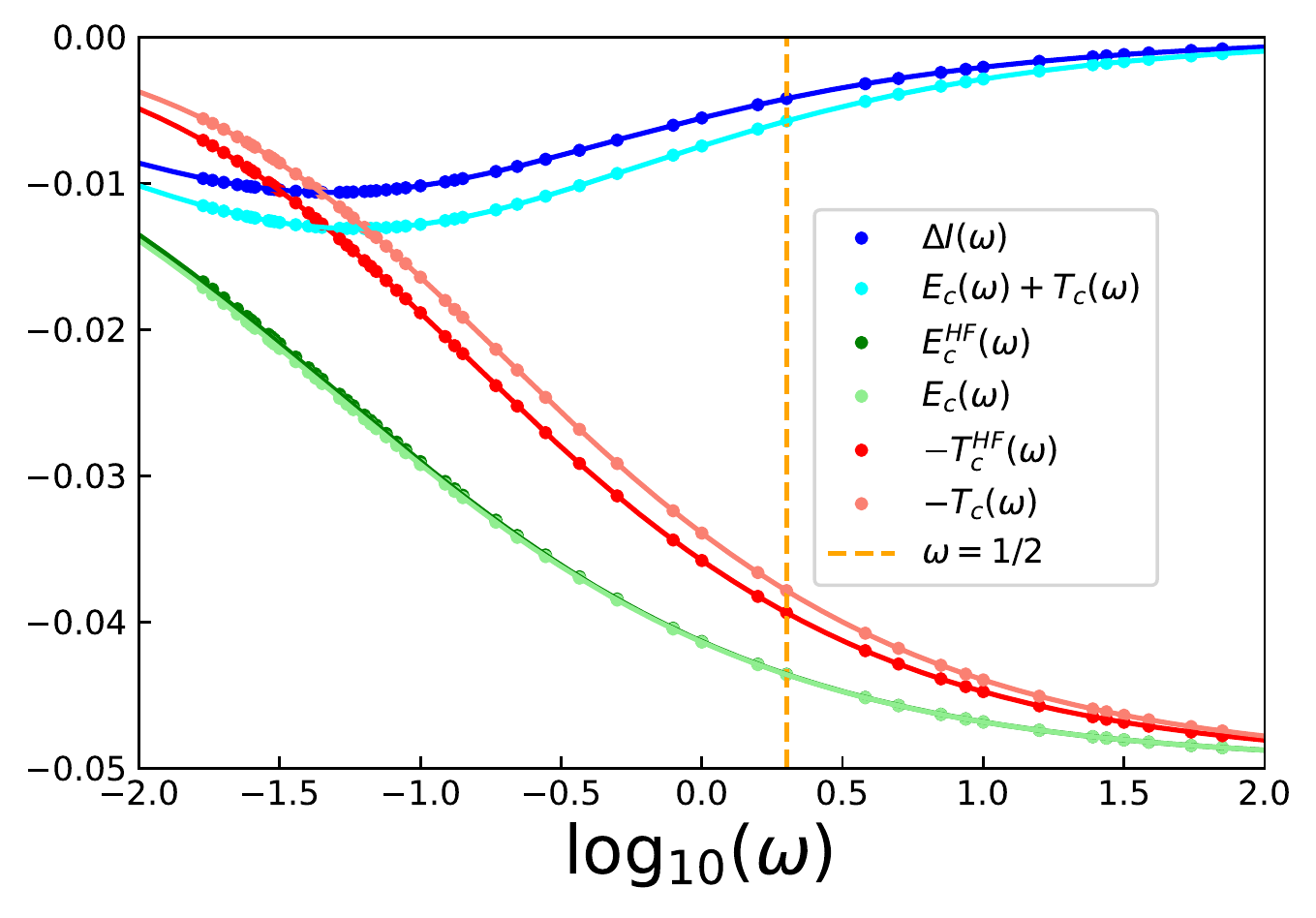}
\end{center}
\caption{ Difference between HF (dark) and DFT (light) components as a function of $\omega$; same scheme as Fig.~(\ref{fig:3}).}
\label{fig:4}
\end{figure}

What follows are plots of the virial quantities for various values of the frequency. The curves change dramatically from small to large $\omega$; in Fig.~\ref{fig:2} we plot against $\log_{10}(\omega)$ to demonstrate overall effectiveness and trends. The fitting functions plotted are piecewise defined, taking the low-density form for $\omega \leq 1/2$ and the high-density form for $\omega > 1/2$. Figure~\ref{fig:3} contains rescaled low and high-density fits to show their effectiveness in each respective limit. We choose $\omega = 1/2$ as the switching point between fits since the energy is known analytically here. 
We also produce the corresponding DFT counterparts and their approximate forms in the high and low-density limits for comparison in Fig.~\ref{fig:4}. For two electrons in spin singlet the non-interacting kinetic energy is given by the Von-Weizsacker kinetic energy density functional \cite{W35} and the exchange energy is simply negative of half the Hartree energy. For $\omega = 1/2$ the exact expression for the kinetic energy can be produced in closed form. Below we detail the errors of the fits in the low and high-density limits in Tables (\ref{tab:coefftab1}-\ref{tab:coefftab3}).

In the low and high-density limits the behavior of the correlation and the kinetic correlation energies are analogous to the conventional values. The fitting functions are taken to be the same as for $E\c\conv$ and $-T\c\conv$, with the only exception being that the coefficients of $-T\c $ were fit separately from $E\c$ and cannot be determined from the Hellmann-Feynman theorem.

\begin{table}

\scalebox{1.2}{
$\begin{array}{|c|c|c|c|c|c|c|}
\hline
\phantom{E\c\conv} & a_{0} & a_{1} & a_{2} & a_{3} & a_{4} & a_{5} \\
\hline
E\c\conv  & 0.618  & 2.190  & 3.900 & 3.956 & 2.197 & 0.518 \\
\hline

E\c   & 0.663  & 2.501  & 4.774 & 5.227 & 3.112 & 0.780 \\
\hline

-T\c\conv  & 0\phantom{.000}  & 1.095 & 3.900 & 5.934 & 4.394 & 1.295 \\
\hline

-T\c   & 0.030  & 1.055 & 3.411 & 4.943 & 3.539 & 1.020 \\
\hline

\end{array}$}
\caption{Coefficients of low density fit Eq.~(\ref{eq:ldf}), in Hartrees }
\label{tab:coefftab1}

\centering

\scalebox{1.2}{
$\begin{array}{|c|c|c|c|c|c|}
\hline
\phantom{E\c\conv} & b_{0} & b_{1} & b_{2} & b_{3} & b_{4} \\
\hline
E\c\conv  & 49.70 &  93.68  & 1.051 & 0.044  & 1.098 \times 10^{-3} \\
\hline

E\c   &  49.70  &  93.68   & 1.114  & 0.069  & 5.819 \times 10^{-3} \\
\hline

-T\c\conv  &  49.70   &  16.40 & 2.628 & 0.141   & 4.395 \times 10^{-3} \\
\hline

-T\c   &  49.70   &  19.34  & 3.944  & 0.038   & 4.306 \times 10^{-3} \\
\hline

\end{array}$}
\caption{Coefficients of high density fit Eq.~(\ref{eq:hdf}), in milliHartrees }
\label{tab:coefftab2}

\scalebox{1.2}{
$\begin{array}{|c|c|c|c|c|c|}
\hline
\phantom{E\c\conv \frac{1}{2}_{+}} & \omega\to \frac{1}{2}^{+} & \omega\to \frac{1}{2}^{-} & \text{Deriv. } \   \\
\hline
E\c\conv  & 2.076 &  -0.063  & 14.73  \\
\hline

E\c   & 2.246  &  -3.333  & 222.1    \\
\hline

-T\c\conv  & 4.845  &  -4.066  & 279.7   \\
\hline

-T\c   &  1.099 &  -0.143  & 261.2  \\
\hline

\Delta E\c   & 0.170 &  -3.270  & 207.4  \\
\hline

\end{array}$}
\caption{Errors of the fitting functions, in microHartrees. The first two columns give the difference between the fit and its respective $\omega = 1/2$ value. The last column gives the difference of the derivatives between the high and low density fits at $\omega =1/2$.}
\label{tab:coefftab3}
\end{table}

\begin{figure}[!htb]
    \includegraphics[width=\linewidth]{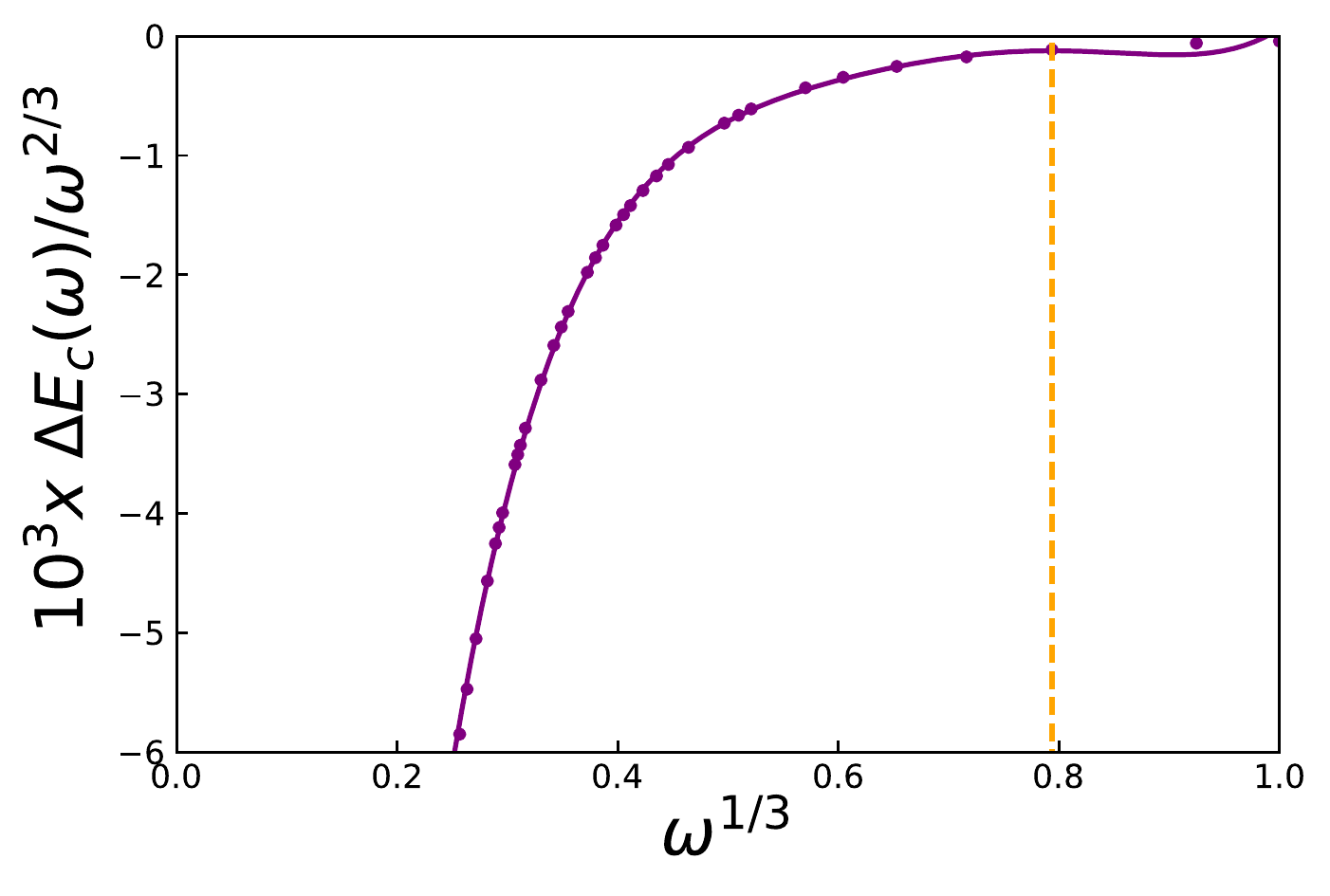}
    \label{fig:5a}  
  \hfill 
    \includegraphics[width=\linewidth]{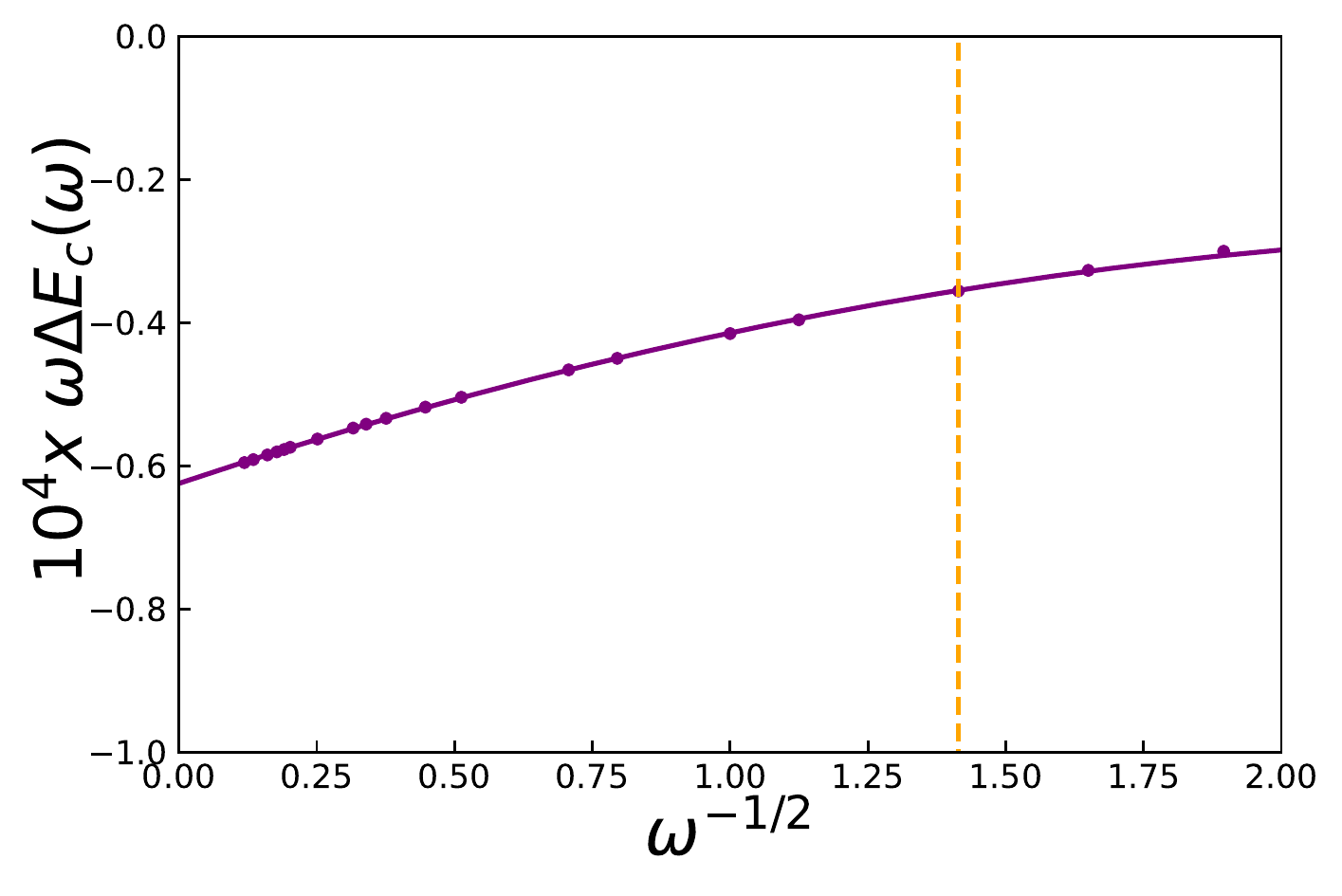}
    \label{fig:5b}
	\caption{Difference between conventional and DFT correlation energies in the low (top) and high (bottom) density limits and their fits, in Hartrees. Dashed orange line denotes $\omega =1/2$.}
	\label{fig:5}
\end{figure}

From Fig.~\ref{fig:5} we find that 
$\Delta E\c  \leq 0$, as we would expect, and this quantity is accurately reproduced by
our fits.
Differences between the high-density fits of $E\c$ and $E\c\conv$ first begin at $\mathcal{O}(\omega^{-1})$. 
For all values of $\omega$ we find that $T \geq T\HF$ from the positivity of the conventional kinetic correlation energy. 
Since we find $T\c\conv > T\c $, the stronger condition $T\s > T\HF$ is satisfied for Hooke's atom.

\sec{Molecules}
\label{mol}

If $N_{A}$ classical nuclei are assumed within the Born-Oppenheimer approximation, the molecular virial integral difference simplifies considerably due to the Hellmann-Feynman theorem,
\ben \Delta I ( \{\textbf{R}_{\alpha}\}) = -\sum_{\alpha = 1}^{N_{A}} \textbf{R}_{\alpha}\cdot \nabla_{\alpha}E\c\conv(\{ \textbf{R}_{\alpha} \} ),   \een
where $E\c\conv(\{ \textbf{R}_{\alpha}\})$ is the conventional correlation energy for an arrangement of nuclei $\{ \textbf{R}_{\alpha}\}$. For this case, the kinetic energy difference can also be cast into a simpler expression,
\ben T\c\conv(\{\textbf{R}_{\alpha}\}) =  \sum_{\beta=1}^{N_{A}} \frac{\partial}{\partial R_{\beta} }\Big(-R_{\beta } E\c\conv(\{\textbf{R}_{\alpha}\}) \Big).\een
We can see that for the case of molecules a sufficient condition to guarantee $T>T\HF$ is the requirement that $-R_{\beta}E\c\conv(\{\textbf{R}_{\alpha} \})$ be monotonically increasing as a function of its radial coordinates $R_{\beta}$.

\begin{figure}[htb]
\centering
\includegraphics[angle=0,width=8.5cm]{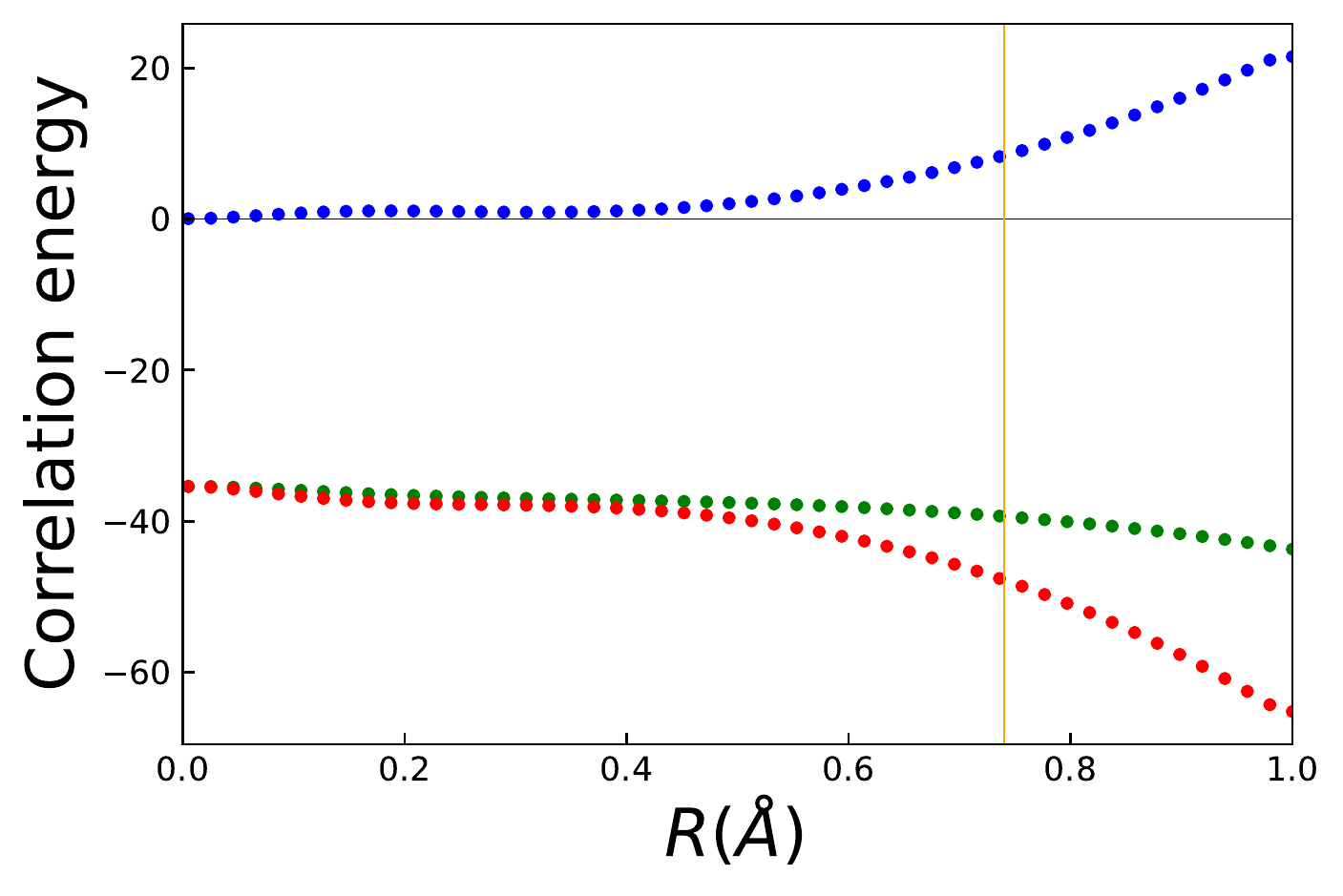}
\caption{$\text{H}_{2}$ conventional correlation energy (green), negative of the conventional
kinetic correlation energy (red), and their difference (blue), in milliHartrees, as a function of nuclear separation. The equilibrium bond-length is given by orange line at $R \approx 0.74$ \AA.  }
\label{fig:0}
\end{figure}

For homonuclear diatomic molecules, the atomic form of the virial theorem holds in both
the united atom limit (zero nuclear separation), and at the dissociation limit (infinite nuclear separation) \cite{S33}.
However, HF will usually break symmetry at sufficiently large separations.  
For classical nuclei in a diatomic molecule, the virial yields
\ben
E\c\conv + T\c\conv  = - R \frac{dE\c\conv}{dR},
\een
where $R$ is the internuclear separation.
Below equilibrium the conventional H$_2$ correlation energy is relatively insensitive to bond-length (at equilibrium, it is -42  \text{mH}, the same
as the united atom limit), we expect $T\c\conv \approx -E\c\conv$ and so remains positive.  
In Fig.~\ref{fig:0} we plot values of the conventional correlation energy and the kinetic energy difference near the
united atom limit of the hydrogen molecule. In the united atom limit, the conventional correlation energy can be seen
to approach a negative constant, where eventually $E\c\conv \approx -T\c\conv$. 
The correlation energies were determined using PySCF \cite{PySCF20}, where the exact values
were computed from CCSD; all calculations were done using cc-pVTZ basis functions. 

\sec{Conclusion}
\label{conc}

We have conjectured that, for cases where a single Slater determinant is the minimizer in a Hartree-Fock calculation,
that the HF kinetic energy is always less than the true kinetic energy, i.e., that of the exact solution of the
Schr\"odinger equation.  We have explored this topic in depth, making use of the virial theorem, but have been
unable to find a proof of this conjecture. We have shown why the Kohn-Sham kinetic energy is always less than the true kinetic energy, by construction, in density function theory.

We numerically calculate several non-trivial examples (the Hooke's atom of two electrons in an oscillator potential
and the H$_2$ molecule away from equilibrium).  In each case, we find no violations of our conjecture.

Our conjecture, if true, might prove to be a useful constraint on wavefunction approximations.   Our conjecture is
limited to Hamiltonians that are in the real-space continuum, and where the inter-electron repulsion is Coulombic.
The case of greatest practical interest is when the external potential is a sum of Coulombic attractions, i.e.,
the non-relativistic, Born-Oppenheimer limit of molecules and solids in the absence of external fields.
We hope the publication of this work will lead to either a proof or a counter-example.

\sec{Acknowledgments}
\label{mel6}
SC and KB were supported by NSF award number CHE-2154371. ML was supported by the Julian Schwinger Foundation.

\bibliography{master.bib}

\end{document}

%% file: Burke_template.tex
\usepackage{amsmath}
\usepackage{physics}
\usepackage{amssymb}
\usepackage{float}
\usepackage[dvipsnames]{xcolor}

\definecolor{TITLECOL}{rgb}{0.05,0.25,0.85}
\definecolor{CONTENTSCOL}{rgb}{0.1,0.2,0.7}
\definecolor{URLCOL}{rgb}{0,0.52,0.83}
\definecolor{LINKCOL}{rgb}{0.05,0.5,0}
\definecolor{CITECOL}{rgb}{0.25,0,0.48}
\definecolor{SECOL}{rgb}{0.07,0.31,0.80}
\definecolor{SSECOL}{rgb}{0.26,0.19,0.75}

\newcommand{\coloredtitle}[1]{\title{\textcolor{TITLECOL}{#1}}}
\newcommand{\coloredauthor}[1]{\author{\textcolor{CITECOL}{#1}}} 
\renewcommand{\sec}[1]{\section{\textcolor{SECOL}{#1}}}

\usepackage{graphicx}

\def\PublicationsLink{http://dft.uci.edu/publications.php}
\def\preprintlink{ \href{\preprintlinklocation}{\TitleOfPaper} }
\def\preprinttext{~}

\usepackage{fancyhdr}
\makeatletter
\fancypagestyle{titlepage}
{	\lhead{\textsc{\preprinttext\ ~ \href{\PublicationsLink}{~}}}
	\chead{}
	\rhead{\preprintlink}
	\lfoot{}}
\makeatother
\def\preprintlink{ 
	\href{\preprintlinklocation}
        {
~}
	}

\definecolor{Green}{rgb}{0.016,0.627,0}
\definecolor{Plum}{rgb}{0.17,0,0.45}
\definecolor{LBlue}{rgb}{0,0.34,0.45}
\definecolor{Sepia}{rgb}{0.37,0.17,0.02}
\definecolor{BurntOrange}{rgb}{0.78,0.39,0}

\usepackage{hyperref}
\hypersetup{
    breaklinks=true,
    bookmarksopen=true,
    bookmarksnumbered=true,
    colorlinks=true,
    linkcolor=LINKCOL,
    linktocpage=true,
    citecolor=CITECOL,
    filecolor=magenta,      
    urlcolor=URLCOL,
}

%% file: KieronsMacros.tex
\def\bea{\begin{eqnarray}}
\def\eea{\end{eqnarray}}
\def\ben{\begin{equation}}
\def\een{\end{equation}}
\def\benu{\begin{enumerate}}
\def\enu{\end{enumerate}}

\def\bei{\begin{itemize}}
\def\eei{\end{itemize}}
\def\beit{\begin{itemize}}
\def\eit{\end{itemize}}
\def\benu{\begin{enumerate}}
\def\enu{\end{enumerate}}

\def\sss{\scriptscriptstyle\rm}

\def\1var{(\bx_1...\bx\N)}

\def\br{{\bf r}}

\def\bx{{x}}

\def\c{_{\sss C}}
\def\s{_{\sss S}}

\def\N{_{\sss N}}

\def\HF{^{\rm HF}}

\def\sph_int{ {\int d^3 r}}
